\documentclass[
   ,final          
  ]
  {aipproc}

\layoutstyle{6x9}
\usepackage{amssymb}
\usepackage{graphicx}
\usepackage{amsmath}

\begin{document}

\title{Condensation and Slow Dynamics of Polar Nanoregions in Lead Relaxors}

\author{D. La-Orauttapong, O. Svitelskiy, and J. Toulouse}
{
address={Physics Department, Lehigh University, Bethlehem, PA 18015}
}

\begin{abstract}
It is now well established that the unique properties of relaxor ferroelectrics are due to the presence of polar nanoregions 
(PNR's). We present recent results from Neutron and Raman scattering of single crystals of PZN, PZN-$x$PT, and PMN. 
Both sets of measurements provide information on the condensation of the PNR's and on their slow dynamics, directly through 
the Central Peak and, indirectly, through their coupling to transverse phonons.  A comparative analysis of these results allows 
identification of three stages in the evolution of the PNR's with decreasing temperature: a purely dynamic stage, a quasi-static 
stage with reorientational motion and a frozen stage.  A model is proposed, based on a prior study of KTN, which 
explains the special behavior of the transverse phonons (TO and TA) in terms of their mutual coupling through the rotations 
of the PNR's.
\end{abstract}

\maketitle
\subsection{Introduction}
For many years, relaxor materials (mostly, lead-based Pb(R$_{1/3}$Nb$_{2/3}$)O$_{3}$, where R = Mg$^{2+}$ or Zn$^{2+}$ 
PMN and PZN respectively), have been a focus of research in the ferroelectrics community. Growing attention has been 
given to their industrially-promising solid solutions with 
PbTiO$_{3}$ (PT)~\cite{Kuwata-etal:1981,Shrout-etal:1990,Park-etal:1997}, especially at 
concentrations near the morphotropic phase boundary (MPB), where their remarkable piezoelectric and electrostrictive 
properties are enhanced~\cite{Park-etal:1997,Liu-etal:1999,Viehland-etal:2001}.  However, the fundamental origin of these 
properties remains a puzzle, and the development of their low-temperature state is still not well understood.

The difficulties in understanding originate from the high complexity of these materials, characterized by chemical, 
compositional and orientational disorder that coexists with the presence of short-range 
order~\cite{Yokomizo-etal:1970,Randall-Bhalla:1990}. As relaxors are 
cooled from high temperature, the major structural changes are preceded by the nucleation of polar nanoregions 
(PNR's)~\cite{Viehland-etal:1992,Burns-Dacol:1983}. These PNR's are the consequence of ion off-centering which is 
commom in many perovskites. By analogy with  KTa$_{1-x}$Nb$_{x}$O$_{3}$ (KTN) at sufficient Nb 
concentrations~\cite{Svitelskiy-Toulouse:2003}, one could expect that their development would lead to the appearance 
of a long-range order. However, unlike KTN where Nb$^{5+}$ is the only off-centered positive ion, lead relaxors 
present a much more difficult case~\cite{Bonneau-etal:1991,Mathan-etal:1991,Verbaere-etal:1992,Uesu-etal:1996,Vakhrushev-etal:1994,
Iwase-etal:1999,Matsushima-etal:2000}. It is often assumed\cite{Husson-etal:1990,Egami-etal:1998,Chen:2000,Blink-etal:2001} 
that Nb$^{5+}$, due to its position in the cell and small radius, acts as the main ferroelectric agent. But, the PNR's in the lead 
compounds, develop in the presence of random fields, that prevent the formation of long-range 
order~\cite{Blink-etal:2001, Krainik-etal:1971}. The frustrating effects of these fields can be compensated by an 
electric field applied in a $\langle111\rangle$ direction~\cite{Ye-Schmid:1993,Mulvihill-etal:1997}. 

The complexity of the lead relaxor materials has resulted in the co-existence of several mutually-excluding interpretations of its light scattering 
spectrum (for review see~\cite{Svitelskiy-Toulouse:2003, Siny-Rev}). Not only is the phonon assignment of particular lines  
in question, but also the very existence of first-order light scattering in a cubic crystal remains unaccounted. 
The absence of a Raman analogy to the "waterfall" 
phenomenon\cite{Gehring-etal:2001,Wakimoto-etal:2002,Naberezhnov-etal:1999,Koo-etal:2002} is also puzzling. First-principle 
calculations of the lattice modes\cite{Prosandeev}, can be very helpful in answering these questions. 

In order to shed some light on the development of the low-temperature phase and to resolve the above mentioned 
contradictions, we decided to carry out a detailed study of the relaxor behavior in 
PZN, PZN-$x$PT ($x$ = 4.5 and 9\%), and PMN single crystals 
(for growth technology see~\cite{Zhang-etal:2000,Chen-etal:2001,Ye-etal:1990}) using neutron and light scattering 
spectroscopy. The purpose of this article is to present a brief summary of the 
work completed~\cite{La-Orauttapong-etal:2001, La-Orauttapong-etal:2003, Svitelskiy-etal:2003, Svitelskiy-Toulouse:2003}. 

\subsection{Neutron scattering studies and their results} 

The neutron scattering experiments were carried out on BT9, HB-1, and 4F-2 triple-axis spectrometers at the NIST Center 
for Neutron Research (NCNR), at the High Flux Isotope Reactor (HFIR) of Oak Ridge National Laboratory, and at the 
Orph\'{e}e reactor of the Laboratoire L\'{e}on Brillouin (LLB), respectively. 
The spectrometer was operated in {\it the final neutron energy $E_f$ fixed} at 14.7 meV ($\lambda = 2.36$~\AA) at NCNR, 
at 13.6 meV ($\lambda = 2.45$~\AA) at HFIR, and in {\it the final neutron wavevector $k_f$ fixed} with either at 1.97 \AA$^{-1}$ 
(8.04 meV, $\lambda$ = 3.19 \AA) or at 1.64 \AA$^{-1}$ (5.57 meV, $\lambda$=3.83 \AA) at LLB. The (002) reflection of a highly 
oriented pyrolytic graphite (HOPG) crystal was used to monochromate and analyze the incident and scattered neutron 
beams. To suppress contamination by higher order neutrons, a HOPG filter was installed in the scattered beam. 
The crystal was mounted onto an aluminum sample holder or a boron nitride and oriented with either in 
the [100]-[011] or [100]-[010] scattering planes. 

Constant-$\vec{Q}$ scans were used to collect data by holding the momentum transfer 
$\vec{Q}$ = $\vec{k}_i$ - $\vec{k}_f$ fixed, while scanning the energy transfer $\hbar\omega$ = $E_i - E_f$. 
Using this scan, the central peak ($\hbar\omega$=0) (CP) and the transverse acoustic (TA) phonon mode were obtained 
upon cooling. 

%
%
\begin{figure}[tbp]
\includegraphics[width=1.\columnwidth]{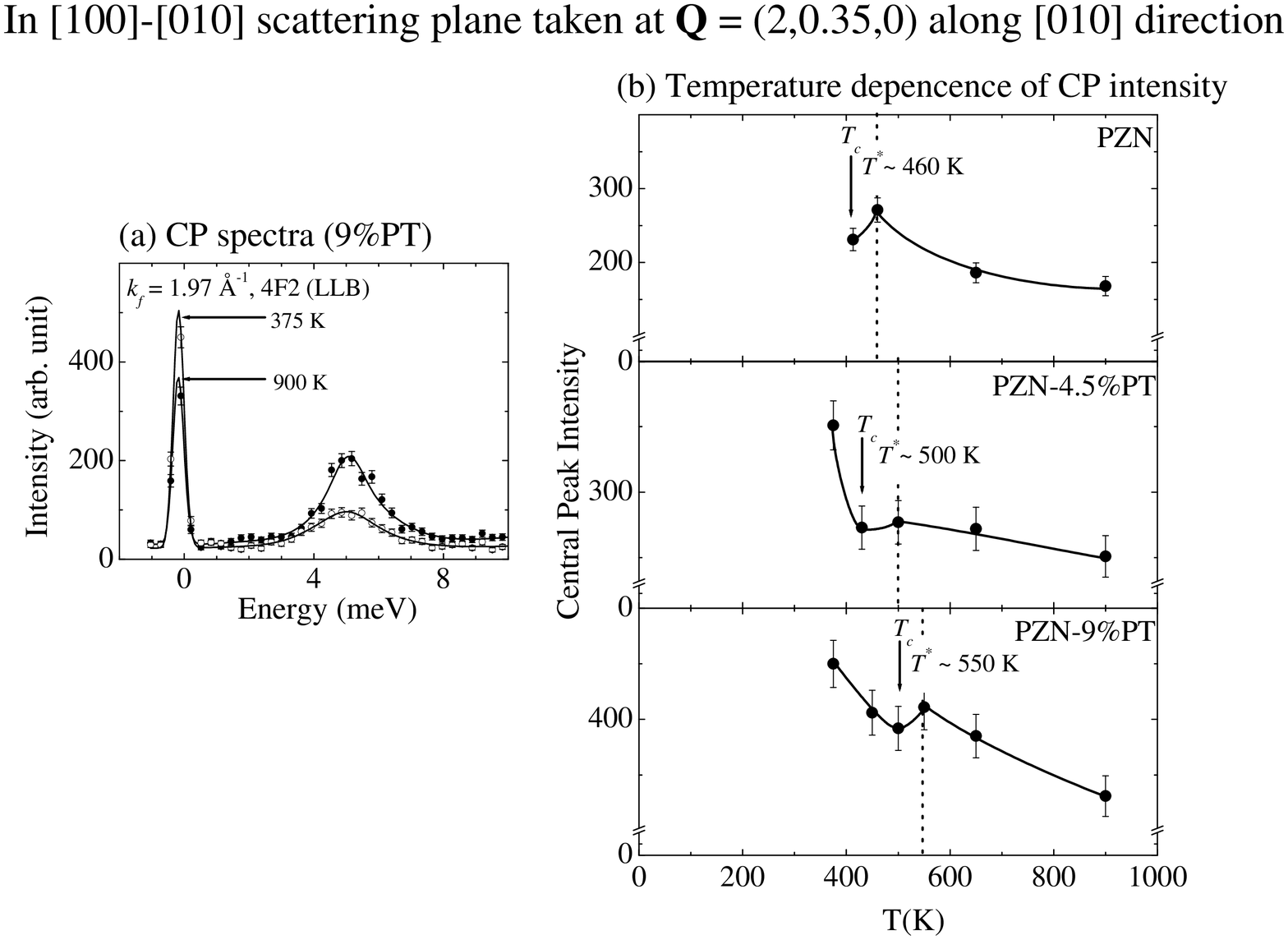}
\caption{(a) Central peak spectra of PZN-9\%PT at $\vec{Q}$ = (2,0.35,0) at 900 K and 375 K
(b) Temperature dependence of the central peak intenisty at $\vec{Q}$ = (2,0.35,0) in PZN, 4.5\%PT,  
and 9\%PT, showing the condensation temperature $T^*$}
\label{fig1}
\end{figure}
%
%

The central peak is a consequence of the relaxational motion in the crystal. So when this motion is fast, the CP is broad and 
its intensity is small. With decreasing temperature, the life time of the clusters increases and their reorientational motion is 
slowing down. This should lead to the growth of the central peak as shown in Fig.~\ref{fig1} (a). This figure show the CP 
spectra are shown taken at $\vec{Q}$ = (2,0.35,0) at 900 K and 375 K ($\vec{Q}=\vec{q}+\vec{G}$, 
where $q$ is the momentum transfer relative to the $\vec{G}$ = (2,0,0) Bragg point, measured along the [010] symmetry 
direction). The CP spectra (solid lines) were fitted to a delta function. 

The temperature dependencies of the CP intensity of 
PZN, 4.5\%, and 9\%PT at $\vec{Q}$ = (2,0.35,0) are presented in Fig.~\ref{fig1} (b). The CP intensity initially 
increases with decreasing temperature until $T \sim T^*$, and goes through a minimum at the transition before increasing 
at lower temperatures. Since the central peak is attributed to the relaxation of the precursor ferroelectric 
clusters\cite{Bruce-Cowley:1981,MichelA}, the presence of a maximum provides supportive evidence for 
a temperature $T^*$ at which the polar regions start to condense. The condensation temperatures 
(PZN : $T^* \sim 460$ K, 4.5\%PT : $T^* \sim 500$ K, and 9\%PT : $T^* \sim 550$ K) found are in agreement 
with our previous neutron {\it elastic} diffuse scattering studies~\cite{La-Orauttapong-etal:2001,La-Orauttapong-etal:2003}. 
It is important to note that, with addition of PT, the PNR's condense at a higher temperature above the transition than in 
pure PZN . This fact is due to stronger correlations between the PNR's in the presence of PT.

%
%
\begin{figure}[tbp]
\includegraphics[width=1.\columnwidth]{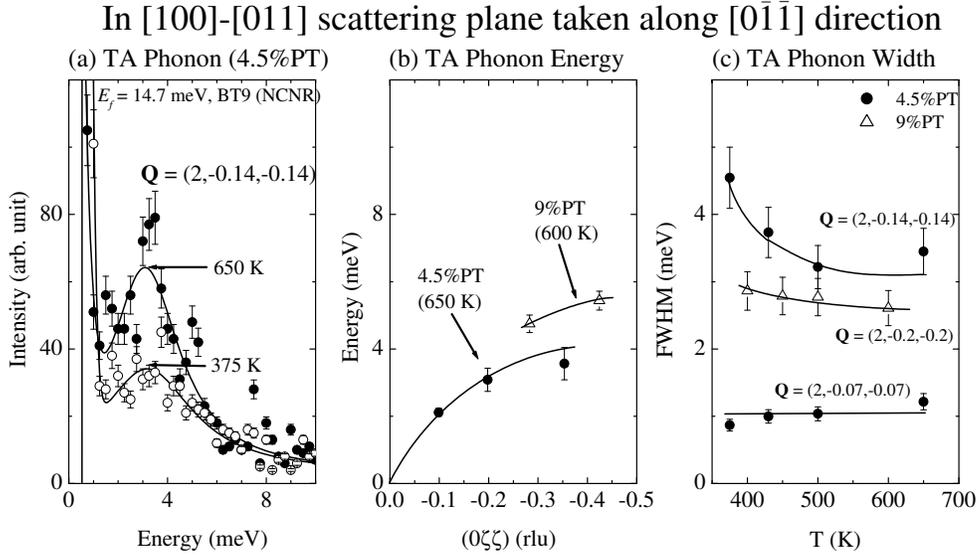}
\caption{(a) TA phonons of 4.5\%PT taken at the scattering vector $\vec{Q}$ = (2,-0.14,-0.14) at 650 K and 375 K
(b) The dispersion curve of the TA mode for 4.5\%PT (650 K) and for 9\%PT (600 K)
(c) Temperature dependence of the FWHM of the TA mode in 4.5\%PT  and 9\%PT}
\label{fig2}
\end{figure}
%
%
%

When investigating the polarization dynamics of relaxors, it is also important to examine the TA phonons, since these should 
couple to the reorientation of the localized strain fields that are known to accompany the polar regions~\cite{Rowe-etal:1979}. 
One might expect to observe increased damping of the TA phonons when the polar regions condense. 
In Fig.~\ref{fig2} (a) we show the TA phonon spectra of 4.5\%PT crystal measured at the scattering vector 
$\vec{Q}$ = (2,-0.14,-0.14) (or $q \sim$ 0.20 rlu)\footnote{1 reciprocal lattice unit (rlu) along 
[0$\bar{1}\bar{1}$] = $\sqrt{2} \times$ rlu = $\sqrt{2}$ ($2\pi/a)$ = 2.19 \AA$^{-1}$, where $a \sim$ 4.05 \AA.} at 650 K and 375 K. 
The TA spectra (solid lines) were fitted to a Lorentzian function. These profiles suggest that the TA mode damping increases with decreasing 
temperature. The peak position ($\hbar\omega$) of the scattered neutron intensity as a function of $|\vec{q}|$ is shown 
in Fig.~\ref{fig2} (b). This figure shows that the frequency of the TA phonon in 9\%PT is higher than in 4.5\%PT. 
However, the width or damping of the phonon is smaller for the 9\%PT than for the 4.5\%PT one as shown in Fig.~\ref{fig2} (c). 
This observation suggests that with increasing PT concentration, the TA damping decreases but the TA phonon frequency 
increases.  

Our measurements of PZN-$x$PT, the TA phonon damping is seen to increase, starting at temperatures far above 
the transition, at a large $q$ first and at a smaller $q$ later with decreasing temperature. In other words, we find that, 
at a given temperature, the larger the $q$, the higher the damping. In fact, we do expect such a trend from the coupling to 
smaller polar regions at higher temperatures and 
to larger and slower ones at lower temperatures. As seen from Fig.~\ref{fig2} (c) for 4.5\%PT, the phonon damping begins to 
increase at $\sim$ 500 K (or $T^{\ast }$) and $\vec{Q}$ = (2,-0.14,-0.14) (or $|\vec{q}| \sim$ 0.20 rlu), which corresponds to about 
5 unit cells ($2\pi/q$). This result is consistent with the size of the polar regions derived from our neutron elastic diffuse 
scattering data~\cite{La-Orauttapong-etal:2003}. Such an agreement provides evidence that the increase in TA phonon 
damping is connected to the appearance of the polar regions. 
In 9\%PT, the TA phonon damping is significant lower than in 4.5\%PT but shows a similar trend.
It is important to note that the measured TA phonon corresponds to the C$_{44}$ elastic modulus, which can couple to 
the reorientations of a strain field with rhombohedral symmetry between different [111] directions. The higher frequency of 
the TA phonon in 9\%PT than in 4.5\%PT indicates that C$_{44}$ is higher in 9\%PT. Both observations, 
lower damping and higher TA phonon frequency, suggest that, with increasing PT, the polar regions are less able to reorient 
and the lattice becomes more rigid. In PMN and PMN-20\%PT\cite{Naberezhnov-etal:1999,Koo-etal:2002}, the TA phonon 
starts to broaden at at temperatures far above the transition or near the wavevector, $q_{wf}$, at which the TO phonon 
has been reported to disappear (``waterfall'')~\cite{Gehring-etal:2001,Wakimoto-etal:2002}.

\subsection{Raman scattering studies and their results}

We have investigated several lead ferroelectric relaxor crystals. In this paper we briefly report on the results obtained on 
PMN $<$100$>$-cut single crystalline sample (for details see \cite{Svitelskiy-etal:2003}). Such a sample represents the 
simplest case and may serve as a model.  

The scattering was excited by propagating in a $\left\langle
100\right\rangle $ direction, 514.5~nm light from a 200~mW Ar$^{+}$-ion
laser, focused to a 0.1~mm spot. The scattered light was collected at an
angle of 90$^{\circ }$ with respect to the incident beam ({\it i.e.}, in $%
\left\langle 010\right\rangle $ direction) by a double-grating spectrometer. For most
of the measurements, the slits were opened to 1.7~cm$^{-1}$. However, in
order to acquire more precise data in the central peak region, at 
temperatures close to the maximum of the dielectric constant ( $100<T<350$~K), 
the slits were narrowed to 0.5~cm$^{-1}$. Each polarization of the
scattered light, $\left\langle x|zz|y\right\rangle $ (VV) and $\left\langle
x|zx|y\right\rangle $ (VH), was measured separately. In order to exclude
differences in sensitivity of the monochromator to different polarizations
of the light, a circular polarizer was used in front of the entrance slit.
For control purposes, we also took measurements without polarization
analysis. Finally, to protect the photomultiplier from the strong Raleigh
scattering, the spectral region from -4 to +4~cm$^{-1}$ was excluded from the
scans. The data were collected in the temperature range from 1000 to 100~K.
The cooling rate was 0.5-1~K/min. Every 50-20~K the temperature was
stabilized and the spectrum recorded. 

The measured spectra were consistent with those
from Refs.\cite{husson} and \cite{ohwa} and shown in \cite{Svitelskiy-etal:2003}.  In the
high temperature region, a typical spectrum consists of two strong lines
centered approximately at 45~cm$^{-1}$ and 780~cm$^{-1}$ 
and of three broad bands between them. The line at 45~cm$^{-1}$ exhibits a triplet structure. 
Lowering temperature leads to the splitting of the broad bands into a number of narrower lines
and to the appearance of new lines. 

To analyze the data, we decomposed the measured spectra using a 
multiple peak fitting procedure. Satisfactory fits could be
achieved with the assumption that the central peak has a Lorentzian shape and that each of the other peaks is described by 
the spectral response function of damped harmonic oscillator, modified by a population factor:

\begin{equation}
\Phi _{i}\sim \frac{\Gamma _{i}f_{0i}^{2}f}{(f^{2}-f_{0i}^{2})^{2}+\Gamma
_{i}^{2}f_{0i}^{2}}F(f,T)\text{ ,}
\end{equation}

\noindent where $\Gamma _{i}$ and $f_{0i}$ are the damping constant and the
mode frequency and the Bose population factor is given by:

\begin{equation}
F(f,T)=\left\{ 
\begin{array}{c}
n(f)+1,\text{ for Stokes part} \\ 
n(f),\text{ for anti-Stokes part}%
\end{array}%
\text{ }\right. \text{, }  \label{boze}
\end{equation}%
where 
\begin{equation*}
n(f)=(exp(hf/kT)-1)^{-1}.
\end{equation*}

As all of the peaks are much better resolved at low temperatures, the fitting procedure was started at the low-temperature 
end of the data set (at 110~K) and, the evolution of the peaks was then followed with increasing
temperature ({\it i.e.}, in the opposite order of the measurements). At
the same time, the number of peaks necessary to achieve
a reasonably good fit was minimized. The control data set (measured without polarization
analysis) was used to calibrate the positions and widths of the weak and poorly resolved
peaks from the VV and VH data sets. Since a large number of parameters is
involved, the results of a particular fit may depend on their initial values. To
stabilize the results, the best-fit values of the parameters obtained at one temperature 
were used as initial values for the fit at the next temperature. In this
manner, several sets of fits were obtained and analyzed. It is remarkable 
that, in all of them, the major parameters showed the same trends. 
Below, we show the most interesting results obtained from the analysis of the central peak and tripet line located at 
45~cm$^{-1}$. For clarity, we describe the observed phenomena from high to low temperatures, following the same order as 
in measurements (unless stated otherwise).

%
%
\begin{figure}[tbp]
\includegraphics[width=1.\columnwidth]{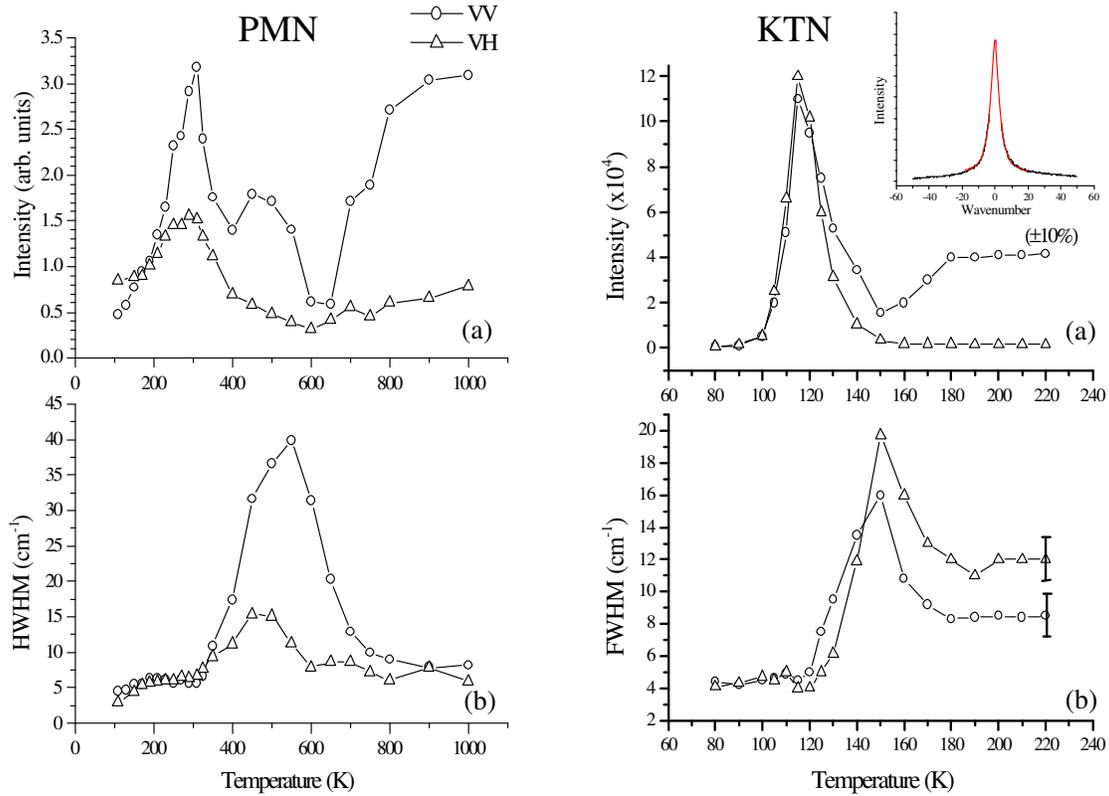}
\caption{Temperature dependencies of the intensity (a) and half-width (b) of
the Lorentzian approximation for the central peak in VV (circles) and VH (triangles) geometries of experiment. Left-hand side 
shows results for PMN crystal, right-hand side compares them with those for KTN crystal \cite{Svitelskiy-Toulouse:2003}. 
Insert demonstrates quality of the fit.}
\label{central}
\end{figure}
%
%

The temperature dependencies of the fitting parameters for the central peak in the PMN crystal
(CP) are presented in Fig. \ref{central} (left-hand side). Circles correspond to the VV and
triangles to the VH component of the peak. The existence of the CP is a direct consequence of the lattice fluctuation relaxations, which 
are very sensitive to the restrictions imposed by the low-symmetry clusters. If the relaxations are fast \cite{MichelA}, the CP is 
low-intense and broad, whereas their slowing causes growth and narrowing of the peak. 

We should point out a striking similarity of the temperature behavior of the CP in PbMg$_{1/3}$Nb$_{2/3}$O$_{3}$ 
(Fig. \ref{central} (left-hand side)) and in KTa$_{0.85}$Nb$_{0.15}$O$_{3}$ (right-hand side) crystals. We have 
shown\cite{Svitelskiy-Toulouse:2003} that the temperature behavior of the CP in KTN can be explained by a model involving 
the collective relaxational motion of off-centered Nb ions and its progressive restriction with decreasing temperature. 
In the cubic phase, Nb ions are allowed to reorient amongst eight equivalent $<$111$>$ directions. The appearance of the 
PNR's, followed by a sequence of phase transitions down to a rhombohedral $R3m$ phase, limits the ion motion to four, two and, 
finally, locks it in only one site. This model is in agreement with the neutron scattering studies of similar systems~\cite{grace}. 
The similarity of the CP behavior in PMN and KTN, suggests that the temperature evolution of the polar 
clusters in PbMg$_{1/3}$Nb$_{2/3}$O$_{3}$ passes through similar stages as those in KTa$_{0.85}$Nb$_{0.15}$O$_{3}$ (KTN). 
However, it is not accompanied by the appearance of long-range order. 

Starting from high temperature, the first important feature is the strong and narrow scattering in the VV geometry, 
accompanied by relatively weak scattering in the VH geometry. This indicates the presence of a symmetric slow 
relaxational motion, most likely involving 180$^\mathrm{o}$ reorientations of ions. Starting from $\sim 900$~K, the cessation 
of this motion causes a decrease in intensity of the CP, which reaches a minimum near the Burns temperature 
$T_{d}\approx620$~K. The prohibition of 180$^\mathrm{o}$ reorientations indicates the loss of inversion symmetry in the 
lattice, caused by onset the distinguishability between Mg and Nb occupied sites and the formation, in the 1:1 ordered areas, 
of a superstructure with average $Fm3m$ symmetry \cite{harmer,boulesteix}. It also imposes the first 
restrictions on the reorientational motion of the dynamical $R3m$ polar nanoregions. Now, they can reorient only amongst 
four neighboring $<$111$>$ directions, forming, on average, tetragonal-like distortions. These processes are accompanied 
by the appearance of large (of the order of wavelength of light) dynamic fluctuations that cause worsening of the optical quality 
of the sample (which is also reflected on the whole spectrum \cite{Svitelskiy-etal:2003}).

With further decrease of the temperature, the optical quality of the crystal improves again. Below $\sim550$~K, the four-site 
reorientational motion of the PNR's slows down, which is marked by the narrowing of the VV component of the CP 
and the increase of its VV and VH intensities. Simultaneously, the VH component broadens and reaches a maximum at 
$\sim450$~K. This indicates rearrangements in the crystalline structure leading to the appearance of new restrictions 
on the ion motion. The analogy with KTN suggests that, below $\sim450$~K, the motion of the $R3m$ clusters becomes restricted 
to two neigboring $<$111$>$ orientations, giving an average monoclinic-like distortion. Such a rearrangement causes some 
decrease in intensity of the VV component (with minimum at $\sim400$~K), while the VH intensity keeps growing. 
Below $\sim400$~K, the slowing down of the two-site relaxational motion causes a narrowing of the CP and an increase in the 
intensities of both components. Below $T_{f}\approx350$~K, these effects become especially dramatic. Further decreasing 
the temperature below $\sim300$~K, leads to the complete prohibition of intersite reorientational motion of the $R3m$ 
clusters, {\it i.e}. to appearance of static $R3m$ clusters. This is marked by a sharp decrease in intensity of both components 
of the CP. At $T_{do}\approx210$~K (which is the temperature of an electric field-induced phase 
transition), the freezing process of the PNR's from dynamic to static is complete. Below this 
temperature, the central peak is narrow and its intensity is small in both scattering geometries.

%
%
\begin{figure}[tbp]
\includegraphics[width=1\columnwidth]{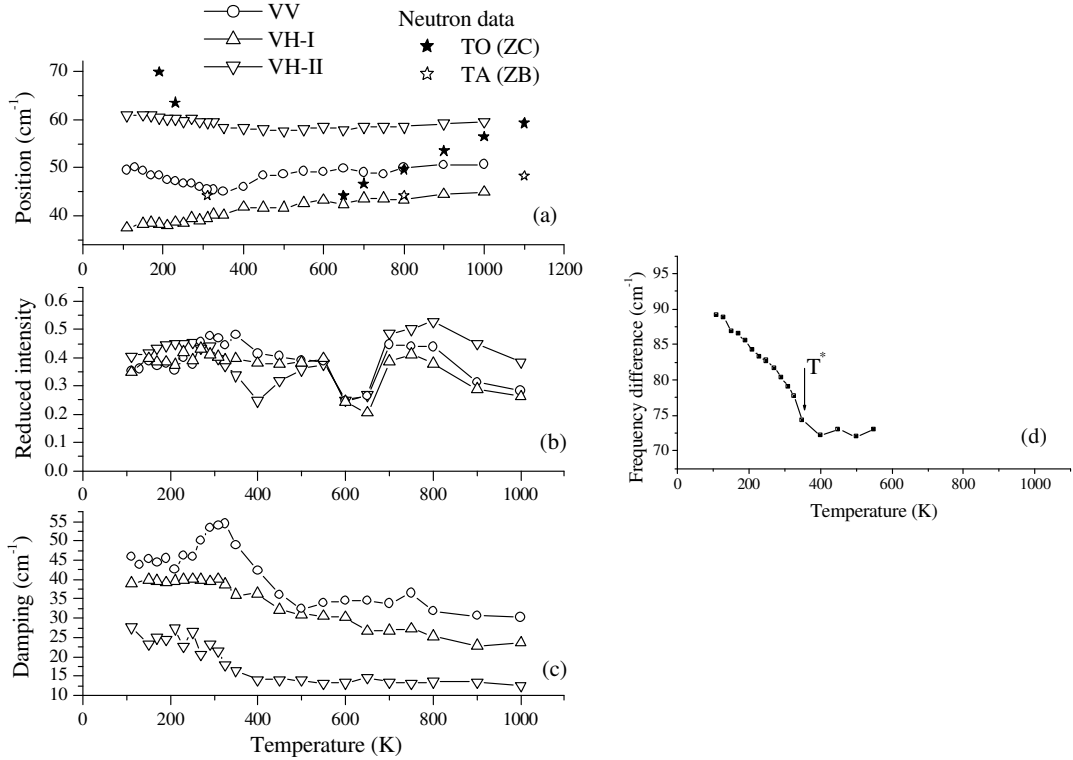}
\caption{(a-c): Temperature dependences of the fitting parameters: position (a),
reduced intensity (b) and damping constant (c) for the triplet line located at 45~cm$^{-1}$. Phonon frequencies, measured by neutron spectroscopy\protect\cite{Gehring-etal:2001, Wakimoto-etal:2002, Naberezhnov-etal:1999, GehVak}, are shown for
comparizon.\newline
(d): Magnitude of splitting between components of the broad band located at 500-600 cm$^{-1}$.}
\label{line45}
\end{figure}
%
%

Figure \ref{line45}(a-c) presents the temperature evolution of the
fitting parameters for the peak located at 45~cm$^{-1}$, showing its position (a), reduced intensity (b)
and damping constant (c). This peak has a triplet structure, containing one component in VV
(circles) and two components in VH (up and down triangles) geometries. The fitting parameters for this peak exhibit changes 
at the above mentioned temperatures $T_{d}$, $T^*$ and $T_{do}$, confirming their significance in the structural evolution 
of the crystal. However, the origin of this peak (see Table I in Ref. \cite{Svitelskiy-etal:2003}) requires clarification. From a 
comparison with the frequencies of the phonon modes measured by neutron spectroscopy, it is clear that this peak cannot 
be due to the zone center soft TO$_{1}$ mode (black stars in Fig. \ref{line45}). On the other hand, the lower frequency VH 
component, and possibly the VV component could be due to disorder-induced scattering TA phonon from the zone boundary (white stars in Fig. \ref{line45}). However, the higher frequency VH component would still not be accounted for.

In an attempt to account for both VH components simultaneously, we have tried to make use of a coupled oscillator 
model~\cite{Svitelskiy-etal:2003}. Results of the fit confirmed the importance of coupling processes in the formation of this line. 
These processes, however, occur without a significant contribution from the zone center TO$_{1}$ phonons, but more likely 
under the influence of the zone boundary TA phonons, with different polarizations propagating in different 
directions~\cite{Prosandeev}. Interaction between such phonons is possible if mediated by relaxational motion of the polar 
nanoregions.

The importance of the temperature $T^*$ is emphasized by the temperature dependence of the splitting between components of the band located at 500-600~cm$^{-1}$ (Fig.\ref{line45}(d)), which is analogous to the one observed in KTN. The work to explain this phenomenon is currently in progress.

\subsection{Conclusions}

By means of neutron and light scattering spectroscopy, we have carried out an investigation of the development of 
the low temperature phase in relaxor ferroelectric materials (PZN-$x$PT and PMN). Our results 
show that the formation of this phase is preceded by appearance of the precursor clusters. These clusters nucleate at very 
high (several hundred degrees higher than the maximum of the dielectric constant) temperatures as highly dynamic objects. 
With lowering of the temperature, their motion becomes progressively more restricted, starting from Burns temperature 
$T_d$. The appearance of the static polar regions is marked by the temperature $T^*$. Finally, the reorientational motion 
freezes out. The process of slowing down is strikingly similar to the one in KTN. However, in the case of lead relaxors, due 
to the presence of frustrating fields, it does not result in the establishment of the long-range order. Phonon-related peaks 
appear in the light scattering spectrum, due to the coupling between phonons of different polarizations, mediated by 
the relaxational motion of the PNR's.

\subsection{Acknowledgements}
This research has been supported by DOE under Contract No. DE-FG02-00ER45842 and by ONR under Grant 
No. N00014-99-1-0738 (Z.-G. Ye).

\end{document}